\documentclass[a4paper,11pt]{article}

\usepackage{jheppub}
\usepackage[T1]{fontenc}
\usepackage{hyperref}
\usepackage{graphicx}
\usepackage{amsmath}
\usepackage{epsfig}
\usepackage{subfig}
\usepackage{xcolor}
\usepackage{natbib}
\usepackage{multirow}

\title{Analysis of double-$J/\psi$ production in $Z$ decay at next-to-leading-order QCD accuracy}
\author{Cong Li}
\author{,}
\author{Zhan Sun}
\author{and}
\author{Gui-Yuan Zhang}
\affiliation{Department of Physics, Guizhou Minzu University, Guiyang 550025, People's Republic of China.}
\emailAdd{sunzhan{\_}hep@163.com}

\abstract{In this article, we study in detail the double-$J/\psi$ yield through $Z$ decay at the next-to-leading-order (NLO) QCD accuracy within the nonrelativistic QCD factorization. At the tree level, the pure QCD diagrams predict a branching ratio of $\mathcal{B}_{Z \to J/\psi+J/\psi} \sim 10^{-12}$; however, the inclusion of the QED diagrams would augment this prediction by approximately 2-3 orders of magnitude. After incorporating the QCD corrections, the QCD results exhibit a considerable increase, whereas the QED results undergo a substantial reduction. Combining the QCD and QED contributions at NLO in $\alpha_s$, it is observed that the prediction of $\mathcal{B}_{Z \to J/\psi+J/\psi}=(1.110^{+0.334+0.054}_{-0.241-0.001})\times 10^{-10}$, which displays a fairly steady dependence on the renormalization scale, is significantly lower than the upper limits released by CMS.}


\begin{document}

\maketitle

\bibliographystyle{JHEP}

\section{Introduction}\label{intro}

In 2019, based on an integrated luminosity of $37.5~\textrm{fb}^{-1}$, the CMS Collaboration conducted the first search for the decay of the $Z$ boson to double $J/\psi$ by detecting their subsequent decay to $\mu^{+}\mu^{-}$ pairs \cite{CMS:2019wch}. The branching ratio is measured to be
\begin{eqnarray}
\mathcal{B}_{Z\to J/\psi+J/\psi} < 2.2 \times 10^{-6} \label{eq1}.
\end{eqnarray}
In 2022, the CMS Collaboration presented the latest measurement of this branching ratio using a larger sample of data corresponding to an integrated luminosity of $138~\textrm{fb}^{-1}$ \cite{CMS:2022fsq}, i.e.
\begin{eqnarray}
\mathcal{B}_{Z\to J/\psi+J/\psi} < 1.4 \times 10^{-6} \label{eq2}.
\end{eqnarray}
The standard-model predictions, calculated using the nonrelativistic QCD (NRQCD) factorization and leading twist light cone model, is of the order of $10^{-12}$ \cite{Likhoded:2017jmx}. These predictions only take into account the contributions of the pure QCD diagrams at the leading-order (LO) accuracy in $\alpha_s$. Very recently, Gao $et~al$. evaluated the QED contributions arising from $Z \to J/\psi+\gamma^{*}$ with $\gamma^{*} \to J/\psi$, suggesting that the virtual-photon effects would notably enhance the QCD results ($\sim 10^{-10}$) \cite{Gao:2022mwa}.

Given the substantial impact of the next-to-leading-order (NLO) QCD corrections to double-charmonium production in $e^{+}e^{-}$ annihilation \cite{{Zhang:2005cha,Gong:2007db,Zhang:2008gp,Brambilla:2010cs,Dong:2011fb,Sun:2018rgx,Sun:2021tma}}, it is necessary to explore whether higher-order terms in $\alpha_s$ could produce a similar significant enhancement in double-$J/\psi$ yield in $Z$ decay, potentially altering the phenomenological implications. To this end, in this paper we will study the $Z$-boson decay into double $J/\psi$ up to the NLO accuracy in $\alpha_s$, incorporating both the QCD and QED diagrams within the NRQCD framework.

The $Z$-boson decay into heavy quarkonium, which has triggered extensive studies \cite{z decay 1,z decay 2,z decay 3,z decay 4,z decay 5,z decay 6,z decay 7,z decay 8,z decay 9,z decay 10,z decay 11,z decay 12,z decay 13,z decay 14,z decay 15,z decay 16,z decay 17,z decay 18,z decay 19,z decay 20,z decay 21,z decay 22,z decay 23,z decay 24,z decay 25,z decay 26,z decay 27,z decay 28,z decay 29,z decay 30,z decay 31,z decay 32,z decay 33,z decay 34,z decay 35,z decay 36,Lansberg:2019adr,Sang:2022erv,Sang:2023hjl}, could offer valuable insights into the mechanism of quarkonium production and provide references for distinguishing between various models. A large number of $Z$ events ($\sim 10^{9}$/year \cite{z decay 25}) can be generated at the LHC, which could be further amplified by the advancements in the HE(L)-LHC upgrades. Furthermore, the proposed CEPC \cite{CEPC} with its clean background will be optimized to achieve the accumulation of approximately $10^{12}$ $Z$-production events within a single operational year. In addition, the subsequent degradation of the $J/\psi$ pair into four muons provides a distinct experimental indication. Based on these perspectives, it appears that obtaining a precise measurement of $\mathcal{B}_{Z \to J/\psi+J/\psi}$, rather than relying on upper limits, holds promise. Our delving into this process beyond the LO accuracy may provide insights into the compatibility of the theoretical predictions with future measurements.

The rest of the paper is organized as follows: Section \ref{cal} is an outline of the calculation formalism. Then, the phenomenological results and discussions are presented in Section \ref{results}. Section \ref{sum} is reserved as a summary.

\section{Calculation formalism}\label{cal}

\subsection{Theoretical Framework}
Within the NRQCD framework \cite{NRQCD1,NRQCD2}, the decay width of $Z \to J/\psi(p_1)+J/\psi(p_2)$ can be factorized as
\begin{eqnarray}
\Gamma=\hat{\Gamma}_{Z \to c\bar{c}[n_1]+c\bar{c}[n_2]}\langle \mathcal{O}^{J/\psi}(n_1)\rangle \langle \mathcal{O}^{J/\psi}(n_2)\rangle,\label{eq3}
\end{eqnarray}
where $\hat{\Gamma}_{Z \to c\bar{c}[n_1]+c\bar{c}[n_2]}$ is the perturbative calculable short distance coefficient, denoting the production of the intermediate state of $c\bar{c}[n_1]$ plus $c\bar{c}[n_2]$. With the restriction to color-singlet contributions, $n_1=n_2=^3S_1^{[1]}$. The universal nonperturbative Long-Distant-Matrix-Element (LDME) $\langle \mathcal{O}^{J/\psi}(n_{1,2})\rangle$ stands for the probabilities of $c\bar{c}[n_{1,2}]$ into $J/\psi$.

The $\hat{\Gamma}_{Z \to c\bar{c}[n_1]+c\bar{c}[n_2]}$ can further be expressed as
\begin{eqnarray}
\hat{\Gamma}_{Z \to c\bar{c}[n_1]+c\bar{c}[n_2]}=\frac{\kappa}{2m_Z}\frac{1}{2!N_s}|\mathcal{M}|^2,\label{eq4}
\end{eqnarray}
where $|\mathcal{M}|^2$ is the squared matrix elements, $1/N_s$ is the spin average factor of the initial $Z$ boson multiplied by the identity factor ($1/2!$) of the two final-state $J/\psi$, and $\kappa$ denotes the factor stemming from the standard two-body phase space. As highlighted in ref. \cite{Gao:2022mwa}, at the LO level in $\alpha_s$, the incorporation of QED diagrams would improve the QCD results by approximately two orders of magnitude. Consequently, our calculations will encompass both QCD and QED diagrams to ensure a comprehensive and accurate estimation.

Built upon the framework used to deal with $e^{+}e^{-} \to J/\psi+\eta_c$ \cite{Sun:2018rgx}, which is similar to $Z \to J/\psi+J/\psi$, we incorporate terms up to the $\alpha^3$ order and achieve NLO accuracy in $\alpha_s$. The squared matrix elements in equation (\ref{eq4}) can then be written as follows,
\begin{eqnarray}
&&\bigg|\left(\mathcal{M}_{\alpha^{\frac{1}{2}}\alpha_{s}}+\mathcal{M}_{\alpha^{\frac{1}{2}}\alpha^2_{s}}\right)+\left(\mathcal{M}_{\alpha^{\frac{3}{2}}}+\mathcal{M}_{\alpha^{\frac{3}{2}}\alpha_{s}}\right)\bigg|^2 \nonumber \\
&=&\big| \mathcal{M}_{\alpha^{\frac{1}{2}}\alpha_{s}} \big|^2+2 \textrm{Re}\left( \mathcal{M}^{*}_{\alpha^{\frac{1}{2}}\alpha_{s}} \mathcal{M}_{\alpha^{\frac{1}{2}}\alpha^2_{s}} \right) \nonumber \\
&& + 2 \textrm{Re}\left( \mathcal{M}^{*}_{\alpha^{\frac{1}{2}}\alpha_{s}} \mathcal{M}_{\alpha^{\frac{3}{2}}} \right)+2 \textrm{Re}\left( \mathcal{M}^{*}_{\alpha^{\frac{1}{2}}\alpha_{s}} \mathcal{M}_{\alpha^{\frac{3}{2}}\alpha_{s}}+\mathcal{M}^{*}_{\alpha^{\frac{3}{2}}} \mathcal{M}_{\alpha^{\frac{1}{2}}\alpha^2_{s}} \right) \nonumber \\
&& + \big| \mathcal{M}_{\alpha^{\frac{3}{2}}} \big|^2+2 \textrm{Re}\left( \mathcal{M}^{*}_{\alpha^{\frac{3}{2}}} \mathcal{M}_{\alpha^{\frac{3}{2}}\alpha_{s}} \right) + \cdots. \label{eq5}
\end{eqnarray}
Correspondingly, we decompose the decay width into three parts,\footnote{The subscripts 1, 2, and 3 represent the $\alpha$ order, while the superscript $0(1)$ denotes the terms of LO(NLO) level in $\alpha_s$.}
\begin{eqnarray}
\Gamma = \Gamma^{(0,1)}_{1}+\Gamma^{(0,1)}_{2}+\Gamma^{(0,1)}_{3},\label{eq6}
\end{eqnarray}
where
\begin{eqnarray}
\Gamma^{(0)}_{1}&\propto&\big| \mathcal{M}_{\alpha^{\frac{1}{2}}\alpha_{s}} \big|^2, \nonumber \\
\Gamma^{(1)}_{1}&\propto&2 \textrm{Re}\left( \mathcal{M}^{*}_{\alpha^{\frac{1}{2}}\alpha_{s}} \mathcal{M}_{\alpha^{\frac{1}{2}}\alpha^2_{s}} \right), \nonumber \\
\Gamma^{(0)}_{2}&\propto&2 \textrm{Re}\left( \mathcal{M}^{*}_{\alpha^{\frac{1}{2}}\alpha_{s}} \mathcal{M}_{\alpha^{\frac{3}{2}}} \right), \nonumber \\
\Gamma^{(1)}_{2}&\propto& 2 \textrm{Re}\left( \mathcal{M}^{*}_{\alpha^{\frac{1}{2}}\alpha_{s}} \mathcal{M}_{\alpha^{\frac{3}{2}}\alpha_{s}}+\mathcal{M}^{*}_{\alpha^{\frac{3}{2}}} \mathcal{M}_{\alpha^{\frac{1}{2}}\alpha^2_{s}} \right), \nonumber \\
\Gamma^{(0)}_{3}&\propto&\big| \mathcal{M}_{\alpha^{\frac{3}{2}}} \big|^2, \nonumber \\
\Gamma^{(1)}_{3}&\propto&2 \textrm{Re}\left( \mathcal{M}^{*}_{\alpha^{\frac{3}{2}}} \mathcal{M}_{\alpha^{\frac{3}{2}}\alpha_{s}} \right).
\end{eqnarray}

\begin{figure}[!h]
\begin{center}
\hspace{0cm}\includegraphics[width=0.65\textwidth]{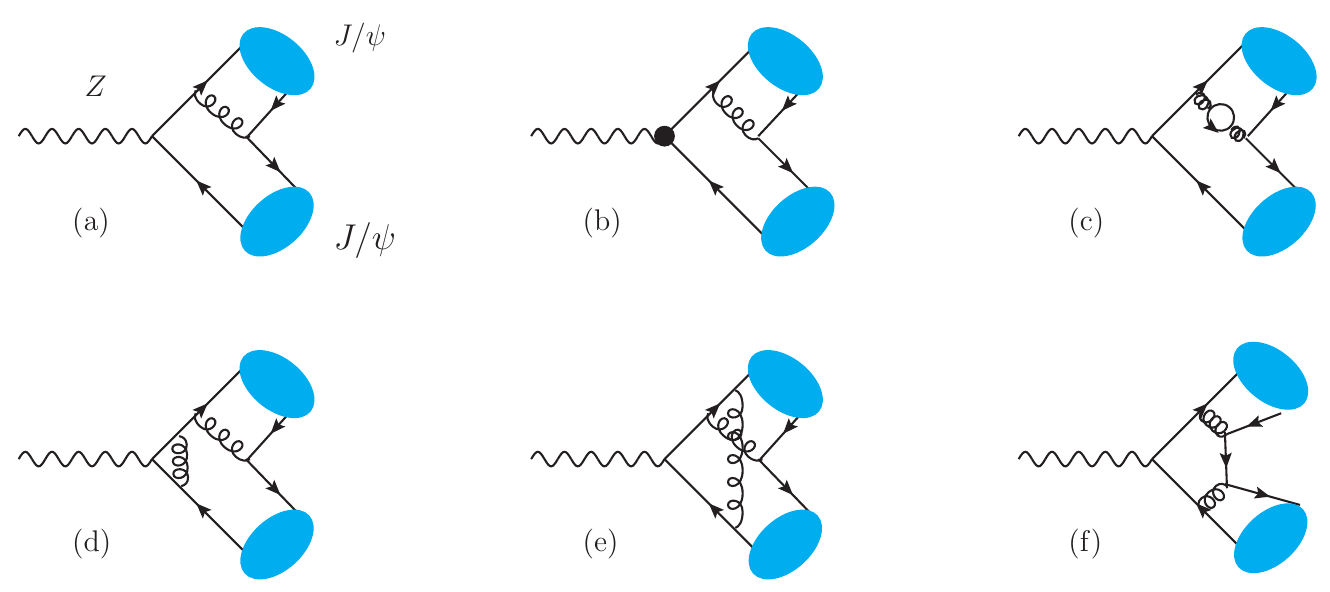}
\caption{\label{fig:QCD Feynman Diagrams}
Representative QCD Feynman diagrams for $Z \to J/\psi(p_1)+J/\psi(p_2)$. $\textbf{a}(\mathcal{M}_{\alpha^{\frac{1}{2}}\alpha_{s}})$ is the tree-level diagram and $\textbf{b}-\textbf{f}(\mathcal{M}_{\alpha^{\frac{1}{2}}\alpha^2_{s}})$ are the NLO QCD corrections to $\textbf{a}$. Diagram $\textbf{b}$ specifically represents the counter-term diagram.}
\end{center}
\end{figure}

\begin{figure}[!h]
\begin{center}
\hspace{0cm}\includegraphics[width=1.0\textwidth]{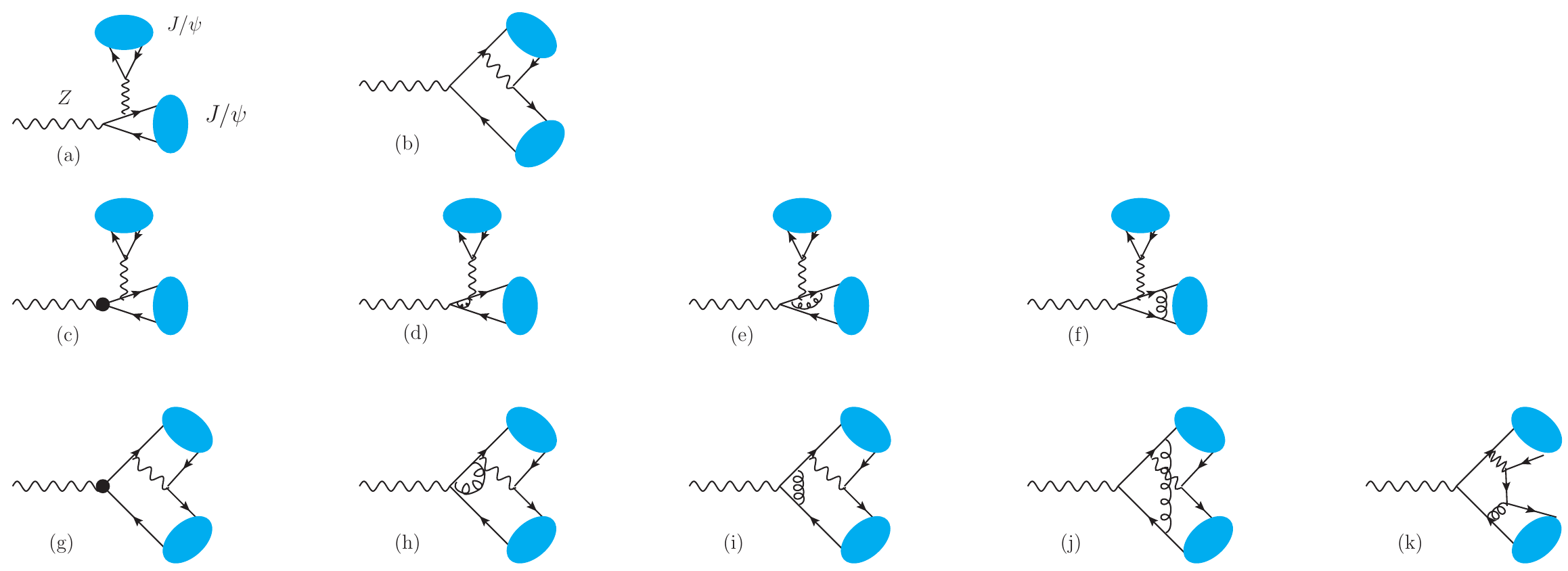}
\caption{\label{fig:QED Feynman Diagrams}
Representative QED Feynman diagrams for $Z \to J/\psi(p_1)+J/\psi(p_2)$. $\textbf{a},\textbf{b}(\mathcal{M}_{\alpha^{\frac{3}{2}}})$ are the tree-level diagram and $\textbf{c}-\textbf{k}(\mathcal{M}_{\alpha^{\frac{3}{2}}\alpha_{s}})$ are the NLO QCD corrections to $\textbf{a},\textbf{b}$. Diagrams $\textbf{c}$ and $\textbf{g}$ specifically represent the counter-term diagram.}
\end{center}
\end{figure}

The representative Feynman diagrams of $\mathcal{M}_{\alpha^{\frac{1}{2}}\alpha_{s}}$, $\mathcal{M}_{\alpha^{\frac{1}{2}}\alpha^2_{s}}$, $\mathcal{M}_{\alpha^{\frac{3}{2}}}$, and $\mathcal{M}_{\alpha^{\frac{3}{2}}\alpha_{s}}$ are displayed in figures \ref{fig:QCD Feynman Diagrams} and \ref{fig:QED Feynman Diagrams}. Figure \ref{fig:QCD Feynman Diagrams}(a) ($\alpha^{\frac{1}{2}}\alpha_{s}$ order) depicts the QCD tree-level diagram (4 diagrams), while figures \ref{fig:QCD Feynman Diagrams}(b-f) illustrate the corresponding NLO QCD corrections (56 one-loop diagrams and 20 counter-term diagrams). Figure \ref{fig:QED Feynman Diagrams}(a,b) ($\alpha^{\frac{3}{2}}$ order) represent the tree-level diagrams for QED (8 diagrams), and figures \ref{fig:QED Feynman Diagrams}(c-k) show the higher-order corrections in $\alpha_s$ (52 one-loop diagrams\footnote{While they exist in figure \ref{fig:QCD Feynman Diagrams}, the one-loop diagrams in figure \ref{fig:QED Feynman Diagrams} do not involve the gluon self-energy and the triple-gluon diagrams.} and 32 counter-term diagrams).

\subsection{The decay width}
The decay width in equation (\ref{eq6}) can generally be written as
\begin{eqnarray}
\Gamma^{\textrm{(0,1)}}_{i}=\frac{\kappa}{2m_Z}\frac{1}{2!N_s} \times \zeta_{i} \times \mathcal{C}^{(0,1)}_{i},~i=1,2,3,
\end{eqnarray}
with
\begin{eqnarray}
\kappa&=&\frac{\sqrt{m^2_Z-16 m^2_c}}{8 \pi m_Z }, \nonumber \\
\zeta_{i}&=&\frac{\pi|R_{1S}(0)|^4}{256 m_c^6 \sin^2\theta_{\textrm{w}} \cos^2\theta_{\textrm{w}}} e^{2(i-1)}_c \alpha^{i}\alpha^{3-i}_s,
\end{eqnarray}
where $N_s=3$ and $e_c=2/3$. The wave function at the origin, denoted as $|R_{1S}(0)|$, can be expressed in terms of NRQCD LDMEs by utilizing the following formulae\footnote{Note that our definition of $\langle \mathcal{O}^{J/\psi}(n_{1,2})\rangle$ differs from that in ref. \cite{NRQCD1} by a factor of $\frac{1}{2N_{\textrm{c}}N_{\textrm{pol}}}$, where $N_{\textrm{pol}}$ represents the number of polarization states of $c\bar{c}[n_{1,2}]$ and $N_{\textrm{c}}$ is equal to 3.}
\begin{eqnarray}
\langle \mathcal O^{J/\psi}(^3S_1^{1}) \rangle = \frac{1}{4\pi}|R_{1S}(0)|^2.
\end{eqnarray}

\subsubsection{LO}

The LO processes depicted in figures \ref{fig:QCD Feynman Diagrams}(a) and \ref{fig:QED Feynman Diagrams}(a,b) are free of divergence, enabling us to straightforwardly determine the coefficients of $\mathcal{C}^{(0)}_{i}$,
\begin{eqnarray}
\mathcal{C}^{(0)}_{1}&=&\frac{65536 m_c^2 ( r^2 -10 r+24)}{9r^4}, \nonumber \\
\mathcal{C}^{(0)}_{2}&=&\frac{16384 m_c^2 (3r+2) ( r^2 -10 r+24)}{3r^4}, \nonumber \\
\mathcal{C}^{(0)}_{3}&=&\frac{1024 m_c^2 (3r+2)^2 ( r^2 -10 r+24)}{r^4},
\end{eqnarray}
where
\begin{eqnarray}
r\equiv\frac{m^2_Z}{4m_c^2}.
\end{eqnarray}
\subsubsection{NLO}

Due to the color conservation, the NLO corrections to $Z \to J/\psi+J/\psi$ do not involve the real correction processes. We utilize the dimensional regularization with $D=4-2\epsilon$ to isolate the ultraviolet (UV) and infrared (IR) divergences. The on-shell mass (OS) scheme is employed to set the renormalization constants for the $c$-quark mass ($Z_m$) and heavy-quark field ($Z_2$); the minimal-subtraction ($\overline{MS}$) scheme is adopted for the QCD-gauge coupling ($Z_g$) and the gluon field $Z_3$. The renormalization constants are taken as
\begin{eqnarray}
\delta Z_{m}^{OS}&=& -3 C_{F} \frac{\alpha_s}{4\pi}N_{\epsilon}\left[\frac{1}{\epsilon_{\textrm{UV}}}+\frac{4}{3}+2\textrm{ln}{2}\right], \nonumber \\
\delta Z_{2}^{OS}&=& - C_{F} \frac{\alpha_s}{4\pi}N_{\epsilon}\left[\frac{1}{\epsilon_{\textrm{UV}}}+\frac{2}{\epsilon_{\textrm{IR}}}+4+6 \textrm{ln}{2}\right], \nonumber \\
\delta Z_{3}^{\overline{MS}}&=& \frac{\alpha_s }{4\pi}(\beta_{0}-2 C_{A})N_{\epsilon}\left[\frac{1}{\epsilon_{\textrm{UV}}}+\textrm{ln}\frac{4m_c^2}{\mu_r^2}\right], \nonumber \\
\delta Z_{g}^{\overline{MS}}&=& -\frac{\beta_{0}}{2}\frac{\alpha_s }{4\pi}N_{\epsilon}\left[\frac{1} {\epsilon_{\textrm{UV}}}+\textrm{ln}\frac{4m_c^2}{\mu_r^2}\right], \label{CT}
\end{eqnarray}
where $N_{\epsilon}= \frac{1}{\Gamma[1-\epsilon]}\left(\frac{4\pi\mu_r^2}{4m_c^2}\right)^{\epsilon}$ is an overall factor, $\gamma_E$ is the Euler's constant, and $\beta_{0}=\frac{11}{3}C_A-\frac{4}{3}T_Fn_f$ is the one-loop coefficient of the $\beta$ function. $n_f(=n_{L}+n_{H})$ represents the number of the active-quark flavors; $n_{L}(=3)$ and $n_{H}(=1)$ denote the number of the light- and heavy-quark flavors, respectively. In ${\rm SU}(3)$, the color factors are given by $T_F=\frac{1}{2}$, $C_F=\frac{4}{3}$, and $C_A=3$.

With the inclusion of the QCD corrections, we acquire the coefficients of $\mathcal{C}^{(1)}_{i}$, which can be expressed in a general form
\begin{eqnarray}
\mathcal{C}^{(1)}_{i}+\mathcal{C}^{(0)}_{i}=\mathcal{C}^{(0)}_{i} \left[ 1+\frac{\alpha_s}{\pi} \left(\xi_{i}\beta_{0}\textrm{ln}\frac{\mu_r^2}{4m_c^2}+a_{i} n_{L}+b_{i} n_{H}+c_{i} \right) \right],\label{NLO exp}
\end{eqnarray}
where $\xi_1=\frac{1}{2}$, $\xi_2=\frac{1}{4}$, and $\xi_3=0$. The coefficients $a_{i}$, $b_{i}$, and $c_{i}$ are dependent solely on the variables of $r$ and $m_c$. One can find their fully-analytical expressions in Appendix \ref{abc1}-\ref{abc3}. We in the following summarize the numerical values assigned to the coefficients of $a_i$, $b_i$, and $c_i$, corresponding to $m_c=1.5 \pm 0.1$ GeV which is often adopted in charmonium-involved processes.

For $m_c=1.4$ GeV,
\begin{eqnarray}
&&a_1=1.3045,~~b_1=1.3026,~~c_1=45.780; \nonumber \\
&&a_2=0.6523,~~b_2=0.6513,~~c_2=20.527; \nonumber \\
&&a_3=0,~~b_3=0,~~c_3=-4.7254.
\end{eqnarray}
For $m_c=1.5$ GeV,
\begin{eqnarray}
&&a_1=1.2586,~~b_1=1.2564,~~c_1=43.832; \nonumber \\
&&a_2=0.6293,~~b_2=0.6282,~~c_2=19.478; \nonumber \\
&&a_3=0,~~b_3=0,~~c_3=-4.8767.
\end{eqnarray}
For $m_c=1.6$ GeV,
\begin{eqnarray}
&&a_1=1.2155,~~b_1=1.2130,~~c_1=42.063; \nonumber \\
&&a_2=0.6078,~~b_2=0.6065,~~c_2=18.523; \nonumber \\
&&a_3=0,~~b_3=0,~~c_3=-5.0183.
\end{eqnarray}

We utilize \texttt{FeynArts} \cite{Hahn:2000kx} to generate all the necessary Feynman diagrams and corresponding analytical amplitudes. Following this, we apply the package \texttt{FeynCalc} \cite{Mertig:1990an} to handle the traces of the $\gamma$ and color matrices, which transforms the hard scattering amplitudes into expressions with loop integrals. When calculating the $D$-dimensional $\gamma$ traces that incorporate a single $\gamma_{5}$ matrix and involve UV and/or IR divergences, following the scheme outlined in refs. \cite{Korner:1991sx,z decay 4, z decay 22}, we choose the same starting point ($Z$-vertex) to write down the amplitudes without implementation of cyclicity. Afterward, we employ our self-written $\textit{Mathematica}$ codes that include implementations of \texttt{Apart} \cite{Feng:2012iq} and \texttt{FIRE} \cite{Smirnov:2008iw} to reduce these loop integrals to a set of irreducible Master Integrals, whose fully-analytical expressions can be found in Appendix \ref{MI1}-\ref{MI11}. As a cross check, we simultaneously adopt the package \texttt{LoopTools} \cite{Hahn:1998yk} to numerically evaluate these Master Integrals, acquiring the same numerical results.

We have refined our calculating framework used in the heavy-quarkonium production in $e^{+}e^{-}$ annihilation \cite{Sun:2018rgx,Sun:2021tma} or $Z$-boson decay \cite{z decay 31,z decay 33,z decay 34,z decay 35} to deal with the calculations within the context. The two processes bear a resemblance of NLO diagrams and $\gamma_{5}$-trace structure to $Z \to J/\psi+J/\psi$. In addition, we have calculated the process of $Z \to J/\psi+\eta_c$ and obtained the same $K$ factor as in the $e^{+}e^{-}$ annihilation.
\section{Phenomenological results}\label{results}
In the calculations, we choose $m_c=m_{J/\psi}/2=1.5 \pm 0.1$ GeV, $m_Z=91.1876$ GeV, $\alpha=1/128$ \cite{z decay 4}, and employ the two-loop $\alpha_s$ running coupling constant. The wave function at the origin is configured as $|R_{1S}(0)|^2=0.81~\textrm{GeV}^3$ \cite{Eichten:1995ch} .

\begin{table*}[htb]
\begin{center}
\caption{Decay widths (in unit: $10^{-12}$ GeV) of $Z \to J/\psi+J/\psi$ with $m_c=1.5$ GeV. ``$\Gamma^{\textrm{LO}}_{\textrm{total}}$" incorporates the individual contributions of $\Gamma^{(0)}_{1}$, $\Gamma^{(0)}_{2}$, and $\Gamma^{(0)}_{3}$, while ``$\Gamma^{\textrm{NLO}}_{\textrm{total}}$" refers to the combined sum of the contributions of $\Gamma^{(0,1)}_{1,2,3}$.}
\label{tab: 1}
\begin{tabular}{|c|cc|cc|cc|cc|cccc}
\hline
$\mu_r$ & $\Gamma^{(0)}_{1}$ & $\Gamma^{(1)}_{1}$ & $\Gamma^{(0)}_{2}$ & $\Gamma^{(1)}_{2}$ & $\Gamma^{(0)}_{3}$ & $\Gamma^{(1)}_{3}$ & $\Gamma^{\textrm{LO}}_{\textrm{total}}$ & $\Gamma^{\textrm{NLO}}_{\textrm{total}}$\\ \hline
$2m_c$ & $1.458$ & $5.850$ & $40.71$ & $73.43$ & $284.2$ & $-115.2$ & $326.4$ & $290.4$\\
$m_Z/2$ & $0.387$ & $1.173$ & $20.98$ & $29.59$ & $284.2$ & $-59.38$ & $305.6$ & $277.0$\\
$m_Z$ & $0.309$ & $0.906$ & $18.76$ & $25.71$ & $284.2$ & $-53.10$ & $303.3$ & $276.8$\\
\hline
\end{tabular}
\end{center}
\end{table*}

\begin{figure}[!h]
\begin{center}
\hspace{0cm}\includegraphics[width=0.55\textwidth]{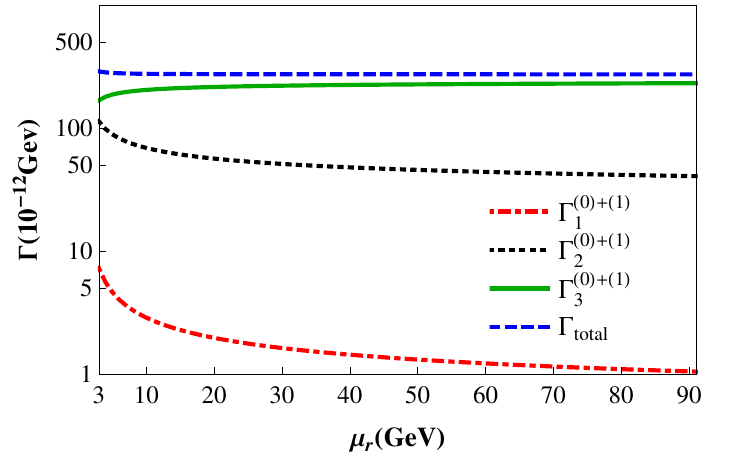}
\caption{\label{fig:mur}
Decay widths of $Z \to J/\psi+J/\psi$ as a function of the renormalization scale with $m_c=1.5$ GeV. ``$\Gamma_{\textrm{total}}$" refers to the combined sum of the contributions of $\Gamma^{(0,1)}_{1,2,3}$.}
\end{center}
\end{figure}

Table \ref{tab: 1} summarizes our predictions of the decay width of $Z\to J/\psi+J/\psi$. Inspecting the data, one can find
\begin{itemize}
\item[1)]
At the LO accuracy in $\alpha_s$, the pure QCD prediction, known as $\Gamma^{(0)}_{1}$, is around $10^{-12}$. However, when the QED diagrams are taken into account, the inclusion of $\Gamma^{(0)}_{2,3}$ significantly enhances the QCD results, resulting in an increase of approximately 2-3 orders of magnitude. The remarkable contribution of the QED diagram can mainly be attributed to the kinematic enhancement arising from the single-photon-fragmentation structure in figure \ref{fig:QED Feynman Diagrams}(a), which compensates sufficiently for the $\alpha$ suppression and then dominates over figure \ref{fig:QCD Feynman Diagrams}(a).
\item[2)]
After incorporating the QCD corrections, the QCD results would experience a significant amplification of approximately 4-5 times, as demonstrated by the ratio of $\left(\Gamma^{(0)}_{1}+\Gamma^{(1)}_{1}\right)\big/\Gamma^{(0)}_{1}$. Of the NLO contributions, the fermion-loop diagram,\footnote{The fermion bubbles include the light quarks ($u,d,s$) and the charm quark.} i.e. figure \ref{fig:QCD Feynman Diagrams}(c), accounts for approximately $20\%$. However, including the higher-order terms in $\alpha_s$ would considerably diminish the pure QED results, e.g., $\left(\Gamma^{(0)}_{3}+\Gamma^{(1)}_{3}\right)\big/\Gamma^{(0)}_{3} \sim 60-80\%$. As a result of the combined influence of enhancement and reduction effects, the QCD corrections will lead to a increase of about 2.5-3 times in the interference terms, as can be verified by referring to $\left(\Gamma^{(0)}_{2}+\Gamma^{(1)}_{2}\right)\big/\Gamma^{(0)}_{2}$. It is worth noting that the rise in $\Gamma^{(1)}_{3}$ towards higher renormalization scale ($\mu_r$) would compensate for the declines in $\Gamma^{(1)}_{1}$ and $\Gamma^{(1)}_{2}$, ultimately causing a steady $\mu_r$ dependence of the total NLO prediction, as depicted in figure \ref{fig:mur}.
\end{itemize}

\begin{table*}[htb]
\begin{center}
\caption{Comparisons of our predicted branching ratio with experiment. ``$\mathcal{B}_{\textrm{theo}}$" stands for the theoretical results, comprised of the contributions of $\Gamma^{(0,1)}_{1,2,3}$. The central value corresponds to $\mu_r=m_Z/2$ and $m_c=1.5$ GeV. The uncertainties in the first column are due to the variation of $m_c$ within the range of $1.4$ to $1.6$ GeV, while the second column of uncertainties are caused by varying $\mu_r$ from $2m_c$ to $m_Z$.}
\label{tab: 2}
\begin{tabular}{|c|c|cc|cc|cc|cccc}
\hline
$\mathcal{B}_{\textrm{exp}}$ \cite{CMS:2022fsq} & ${\mathcal{B}}_\textrm{theo}$ \\ \hline
$<1.4\times10^{-6}$ & $(1.110^{+0.334+0.054}_{-0.241-0.001})\times 10^{-10}$\\
\hline
\end{tabular}
\end{center}
\end{table*}

At last, we validate our predictions with experimentation. The results, as presented in table \ref{tab: 2}, indicate that the calculated $\mathcal{B}_{Z \to J/\psi+J/\psi} $ noticeably undershoot the upper limit established by the CMS Collaboration. The variation of $m_c$ by 0.1 GeV around 1.5 GeV would lead to a prediction alteration by $20-30\%$. Conversely, the deviation of $\mu_r$ from $2m_c$ to $m_Z$ around $m_Z/2$ just has a slight impact on the predictions. The upper experimental bound instead of precise measurement implies the failure to detect this branching above the existing $Z$ production rate at the currently running LHC. Hopefully, the future CEPC experiment will have the capability to provide a detailed measurement of this decay channel.

\section{Summary}\label{sum}

In order to provide deeper insights into $Z \to J/\psi+J/\psi$, we in this paper conducted a NLO study of this process using the NRQCD framework. Our LO results indicate that the contributions from QED diagrams dominate over those from QCD diagrams. The QCD corrections have a substantial amplifying effect on the QCD results, while simultaneously diminishing the QED results. By considering both the QCD and QED contributions, we estimated the branching ratio to be $\mathcal{B}_{Z \to J/\psi+J/\psi}=(1.110^{+0.334+0.054}_{-0.241-0.001})\times 10^{-10}$, which exhibits a rather steady renormalization-scale dependence. This prediction falls significantly below the upper limits set by the CMS Collaboration.

\appendix
\section{Analytical NLO expressions}
In this section, we list the analytical expressions of the coefficients $a_{i}$, $b_{i}$, and $c_{i}$ in equation (\ref{NLO exp}), which are written as a superposition of the Master Integrals $\mathcal{I}$.
\subsection{NLO coefficients}
\begin{eqnarray}
a_1&=&\frac{-3\mathcal I^{(2)}_2+1-6\ln(2m_c)}{9}, \nonumber
\end{eqnarray}
\begin{eqnarray}
b_1&=&\frac{6\mathcal I_{1}-3m_c^2 (r+2)\mathcal I^{(3)}_2+m_c^2\left[r-6-6r\ln(2m_c)\right]}{9m_c^2r}, \nonumber
\end{eqnarray}
\begin{eqnarray}
c_1&=&\frac{76r^4 -151r^3 +228r^2 +216r+24}{3m_c^2 r (2r+1)(r^2 -10r+24)}\mathcal I_{1}+\frac{-496r^4 +109r^3 +1000r^2 +644 r+120}{6r^4 -57r^3+114r^2+72r}\mathcal I^{(1)}_2 \nonumber \\
&&+\frac{179r^2 -469 r+348}{3r^2 -30r+72}\mathcal I^{(2)}_2+\frac{-5r^2 -68 r+304}{18(r^2 -10r+24)}\mathcal I^{(3)}_2+\frac{2(95r^2 +131 r-412)}{9(r^2 -10r+24)}I^{(4)}_2\nonumber \\
&&+\frac{4m_c^2 r(r^2 -5 r+20)}{3(r^2 -10r+24)}\mathcal I^{(1)}_3-\frac{m_c^2 (113r^3+160r^2 -712 r+240)}{24(r^2 -10r+24)}\mathcal I^{(2)}_3+\frac{m_c^2 (7r^3-52r^2 +50 r+48)}{3(r^2 -10r+24)}\mathcal I^{(3)}_3\nonumber \\
&&+\frac{m_c^2 (-303r^3+144r^2 +488r-144)}{24(r^2 -10r+24)}\mathcal I^{(4)}_3+\frac{2m_c^2(11r^3 +66r^2 -198 r+120)}{3(r^2 -10r+24)}\mathcal I^{(5)}_3\nonumber \\
&&+\frac{2m_c^2(r^3 -6r^2 +24 r-48)}{3(r^2 -10r+24)}\mathcal I^{(6)}_3-\frac{179r^3-3146r^2 +6936 r+1584}{18r(r^2 -10r+24)}+(\frac{4}{r}+21)\ln(m_c)+11\ln(2). \nonumber \\
\label{abc1}
\end{eqnarray}
\begin{eqnarray}
a_2&=&\frac{-3\mathcal I^{(2)}_2+1-6\ln(2m_c)}{18}, \nonumber
\end{eqnarray}
\begin{eqnarray}
b_2&=&\frac{6\mathcal I_{1}-3m_c^2 (r+2)\mathcal I^{(3)}_2+m_c^2\left[r-6-6r\ln(2m_c)\right]}{18m_c^2r}, \nonumber
\end{eqnarray}
\begin{eqnarray}
c_2&=&\frac{177r^5 -484r^4+658r^3 +792r^2 +720r+264}{3m_c^2 r (2r+1)(3r+2)(r^2 -10r+24)}\mathcal I_{1}\nonumber \\
&&+\frac{-1500r^5-2761r^4 +5650r^3 +7264r^2 +2000 r+48}{6r(6r^4 -53r^3+76r^2+148r+48)}\mathcal I^{(1)}_2+\frac{537r^3-769r^2 -790 r+1464}{6(3r+2)(r^2 -10r+24)}\mathcal I^{(2)}_2 \nonumber \\
&&-\frac{15r^3+134r^2 -1864 r+4256}{36(3r+2)(r^2 -10r+24)}\mathcal I^{(3)}_2+\frac{267r^3+1655r^2 -2566 r-280}{9(3r+2)(r^2 -10r+24)}I^{(4)}_2\nonumber \\
&&+\frac{2m_c^2 r(3r^3-56r^2 +184 r-64)}{3(3r+2)(r^2 -10r+24)}\mathcal I^{(1)}_3-\frac{m_c^2 (339r^4+1058r^3-3512r^2 +3776 r-3360)}{48(3r+2)(r^2 -10r+24)}\mathcal I^{(2)}_3\nonumber \\
&&+\frac{m_c^2 (21r^4-110r^3-238r^2 +1028 r-672)}{6(3r+2)(r^2 -10r+24)}\mathcal I^{(3)}_3-\frac{m_c^2 (909r^4+654r^3-2904r^2 +1504r-2016)}{48(3r+2)(r^2 -10r+24)}\mathcal I^{(4)}_3\nonumber \\
&&+\frac{m_c^2(21r^4+506r^3-1050r^2 +252 r+48)}{3(3r+2)(r^2 -10r+24)}\mathcal I^{(5)}_3+\frac{m_c^2(3r^4-32r^3 +156r^2 -480 r+672)}{3(3r+2)(r^2 -10r+24)}\mathcal I^{(6)}_3\nonumber\\
&&-\frac{1311r^4-17284r^3+28324r^2 +39984 r+3744}{36r(3r+2)(r^2 -10r+24)}+(\frac{4}{r}+\frac{31}{2})\ln(m_c)+\frac{11\ln(2)}{2}. \label{abc2}
\end{eqnarray}
\begin{eqnarray}
a_3&=&b_3=0, \nonumber 
\end{eqnarray}
\begin{eqnarray}
c_3&=&\frac{126r^5-667r^4 +934r^3 +480r^2 +936r+480}{3m_c^2 r (2r+1)(3r+2)(r^2 -10r+24)}\mathcal I_{1}-\frac{4(3r^5+524r^4 -608r^3 -833r^2 -88 r+48)}{3 r (2r+1)(3r+2)(r^2 -10r+24)}\mathcal I^{(1)}_2 \nonumber \\
&&+\frac{8(35r^2 -112 r+96)}{3(3r+2)(r^2 -10r+24)}\mathcal I^{(2)}_2+\frac{8(5r^2 +68 r-304)}{9(3r+2)(r^2 -10r+24)}\mathcal I^{(3)}_2-\frac{4(9r^3-536r^2 +796 r-272)}{9(3r+2)(r^2 -10r+24)}I^{(4)}_2\nonumber \\
&&-\frac{4m_c^2 r(43r^2 -134 r+104)}{3(3r+2)(r^2 -10r+24)}\mathcal I^{(1)}_3+\frac{4m_c^2 (-11r^3+53r^2 -140 r+120)}{3(3r+2)(r^2 -10r+24)}\mathcal I^{(2)}_3\nonumber \\
&&+\frac{4m_c^2 (8r^3-71r^2 +196 r-192)}{3(3r+2)(r^2 -10r+24)}\mathcal I^{(3)}_3-\frac{4m_c^2 (15r^3-36r^2 +64r-72)}{3(3r+2)(r^2 -10r+24)}\mathcal I^{(4)}_3\nonumber \\
&&-\frac{4m_c^2(6r^4-143r^3 +294r^2 -144 r+96)}{3(3r+2)(r^2 -10r+24)}\mathcal I^{(5)}_3-\frac{32m_c^2(r^3 -6r^2 +24 r-48)}{3(3r+2)(r^2 -10r+24)}\mathcal I^{(6)}_3\nonumber \\
&&-\frac{387r^4-4102r^3+6904r^2 +10680 r+288}{9r(3r+2)(r^2 -10r+24)}+(\frac{4}{r}+10)\ln(m_c).\label{abc3}
\end{eqnarray}
\subsection{Master Integrals}
Here we just present the finite ($\epsilon^0$-order) terms of the Master Integrals in \ref{abc1}-\ref{abc3}. For brevity, we define
\begin{eqnarray}
&&a=\sqrt{r},~b=\sqrt{r -1},~c=\sqrt{r -4},~d=2 r+1, \nonumber \\ &&f=r+ac-4,~g=r-ac-4,~h=r g+2ac, \nonumber\\
&&j=r f-2ac,~j_1=(r-4) a b,~j_2=(2-r)b c.
\end{eqnarray}
There is only one 1-point scalar integral
\begin{eqnarray}
\mathcal I_{1}&=&\frac{\lambda}{\mu_r^{4-D}}\int{\frac{d^D k}{k^2-m_c^2}}=m_c^2 \left[1-2\ln(m_c)\right], \label{MI1}
\end{eqnarray}
where $k$ denotes the loop momentum, $\lambda=\frac{\mu_r^{4-D}}{i\pi^{\frac{D}{2}}\gamma_\Gamma}$ with $\gamma_\Gamma = \frac{\Gamma^2 (1-\epsilon)\Gamma (1+\epsilon)}{\Gamma (1-2\epsilon)}$.\\
There are four 2-point scalar integrals,
\begin{eqnarray}
\mathcal I^{(1)}_2&=&\frac{\lambda}{\mu_r^{4-D}}\int{\frac{d^D k}{k^2[(k+\frac{2p_1+p_2}{2})^2-m_c^2]}}=2-2\ln{(m_c)}-\frac{2r}{d} \left[\ln{(2r)}-i\pi\right], \nonumber \\ \label{MI2}
\end{eqnarray}
\begin{eqnarray}
\mathcal I^{(2)}_2&=&\frac{\lambda}{\mu_r^{4-D}}\int{\frac{d^D k}{k^2(k-\frac{p_1+p_2}{2})^2}}=2-\ln{({m_c}^2 r)}+i\pi. \label{MI3}
\end{eqnarray}
\begin{eqnarray}
\mathcal I^{(3)}_2&=&\frac{\lambda}{\mu_r^{4-D}}\int \frac{d^D k}{(k^2-m_c^2)[(k+\frac{p_1+p_2}{2})^2-m_c^2]}=\frac{c}{a}\left[\ln{ \left(\frac{4r}{(f+4)^2}\right)}+i\pi\right]+2\left[1-\ln{(m_c)}\right],\nonumber \\ \label{MI4}
\end{eqnarray}
\begin{eqnarray}
\mathcal I^{(4)}_2&=&\frac{\lambda}{\mu_r^{4-D}}\int\frac{d^D k}{(k^2-m_c^2)[(k+p_1+p_2)^2-m_c^2]} =\frac{b}{a}\left[\ln{(-2ab+d-2)}+i\pi\right]+2\left[1-\ln{(m_c)}\right].\nonumber \\ \label{MI5} 
\end{eqnarray}
There are six 3-point scalar integrals,
\begin{eqnarray}
\mathcal I^{(1)}_3&=&\frac{\lambda}{\mu_r^{4-D}}\int \frac{d^D k}{k^2[(k+\frac{p_2}{2})^2-m_c^2][(k+\frac{2p_1+p_2}{2})^2-m_c^2]} \nonumber\\
&&=\frac{1}{2 a c m_c^2}\left\{{\ln(2)\ln\left(\frac{(3(g+4)+h)^2}{d(g+h+4)^2}\right)+\left[\ln(r)-i\pi\right]\ln\left(\frac{d}{(g+3)^2}\right)+4 \textrm{Li}_2 \left(\frac{c}{a}\right)+\textrm{Li}_2 \left(-\frac{2(ac+j)}{d}\right)}\right.\nonumber\\
&&\left.{-\textrm{Li}_2 \left(\frac{c^2}{ac-h}\right)-\textrm{Li}_2 \left(-\frac{c+ag}{a}\right)-2\textrm{Li}_2 \left(-\frac{c}{a}\right)}\right\}, \label{MI6}
\end{eqnarray}
\begin{eqnarray}
\mathcal I^{(2)}_3&=&\frac{\lambda}{\mu_r^{4-D}}\int \frac{d^D k}{k^2[(k-\frac{p_2}{2})^2-m_c^2](k+\frac{p_1+p_2}{2})^2}\nonumber\\
&&=\frac{1}{ a c m_c^2}\left\lbrace{\ln(2)\ln\left(\frac{d(f+4)^2}{(3r-ac)^2}\right)+\left[\ln(r)-i\pi)\right]\ln\left(\frac{4dr}{(3r-ac)^2}\right)+\textrm{Li}_2\left(-\frac{2(ac+j)}{d}\right)+2\textrm{Li}_2 \left(\frac{c}{a}\right)}\right. \nonumber \\
&&\left.{+\textrm{Li}_2 \left(-\frac{g}{2}\right)-\textrm{Li}_2 \left(-\frac{f}{2}\right)-\textrm{Li}_2 \left(\frac{c^2}{ac-h}\right)-\textrm{Li}_2  \left(-\frac{c+ag}{a}\right)}\right\rbrace, \label{MI7}
\end{eqnarray}
\begin{eqnarray}
\mathcal I^{(3)}_3&=&\frac{\lambda}{\mu_r^{4-D}}\int \frac{d^D k}{k^2[(k+\frac{p_1}{2})^2-m_c^2](k+\frac{p_1+p_2}{2})^2} \nonumber\\
&&= \frac{1}{ a c m_c^2}\left\lbrace \left[\ln(r)-i\pi\right]\ln\left(\frac{(f+4)^2}{4r}\right)+2\textrm{Li}_2 \left(\frac{c}{a}\right)-2\textrm{Li}_2 \left(-\frac{c}{a}\right)-\textrm{Li}_2 \left(-\frac{g}{2}\right)+\textrm{Li}_2 \left(-\frac{f}{2}\right)\right\rbrace, \nonumber \\  \label{MI8}
\end{eqnarray}
\begin{eqnarray}
\mathcal I^{(4)}_3&=&\frac{\lambda}{\mu_r^{4-D}}\int \frac{d^D k}{k^2[(k+\frac{2p_1+p_2}{2})^2-m_c^2](k+\frac{p_1+p_2}{2})^2} \nonumber\\
&&= \frac{1}{ a c m_c^2}\left\lbrace{\ln(2)\ln\left(\frac{d}{(g+3)^2}\right)+\left[\ln(r)-i\pi)\right]\ln\left(\frac{g+6}{2g+6}\right)+\textrm{Li}_2 \left(-\frac{g}{2}\right)-\textrm{Li}_2 \left(-\frac{f}{2}\right)-2\textrm{Li}_2 \left(-\frac{c}{a}\right)}\right. \nonumber \\
&&\left.{-\textrm{Li}_2 \left(\frac{2ac-2h}{d}\right)+\textrm{Li}_2 \left(\frac{h-ac}{dr}\right)+\textrm{Li}_2 \left(\frac{c}{a}-f\right)}\right\rbrace, \label{MI9}
\end{eqnarray}
\begin{eqnarray}
\mathcal I^{(5)}_3&=&\frac{\lambda}{\mu_r^{4-D}}\int \frac{d^D k}{k^2[(k-\frac{p_1}{2})^2-m_c^2][(k+\frac{p_1+2p_2}{2})^2-m_c^2]}\nonumber\\
&& = \frac{1}{2 a c m_c^2}\left\{{\left[\ln\left(\frac{(r-ab)^2}{r}\right)+i\pi\right]\ln\left(\frac{br(2b-3c)+j-j_1}{br(2b+3c)+j+j_1}\right)+\left[\ln(2r)-i\pi\right]\ln\left(\frac{(g+3)^2}{d}\right)}\right. \nonumber \\
&&\left.{{-\textrm{Li}_2 \left(-\frac{2(ac+j)}{d}\right)-\textrm{Li}_2 \left(\frac{c^2 -h+j_1+j_2}{r}\right)-\textrm{Li}_2 \left(-\frac{-c^2 +h+j_1+j_2}{r}\right)+\textrm{Li}_2 \left(\frac{c^2 -j+j_1-j_2}{r}\right)}}\right.\nonumber \\
&&\left.{-2\textrm{Li}_2 \left(-\frac{c}{a}\right)+\textrm{Li}_2 \left(\frac{c^2 -j-j_1+j_2}{r}\right)+{\textrm{Li}_2 \left(\frac{c^2}{ac-h}\right)+\textrm{Li}_2 \left(-\frac{c+ag}{a}\right)}}\right\}, \nonumber \\ \label{MI10}
\end{eqnarray}
\begin{eqnarray}
\mathcal I^{(6)}_3&=&\frac{\lambda}{\mu_r^{4-D}}\int \frac{d^D k}{k^2[(k+\frac{p_2}{2})^2-m_c^2][(k+\frac{p_1+2p_2}{2})^2-m_c^2]} \nonumber\\
&&= \frac{1}{ a c m_c^2}\left\{{\ln(2)\ln \left( \frac{(f+4)^4\left(3(2r-ac)+h\right)}{4(g+6)\left(4(r-ac)+h\right)}\right)+\ln(r)\ln\left(\frac{dr}{(d+f+3)^2}\right)-i\pi\ln\left(\frac{16d}{(g+6)^2 r}\right)}\right.\nonumber \\
&&\left.{+\textrm{Li}_2 \left(-\frac{g}{8}\right)+\textrm{Li}_2 \left(-\frac{4ac+j}{2d}\right)-\textrm{Li}_2 \left(\frac{4ac-h}{4r}\right)+2\textrm{Li}_2 \left(\frac{c}{2a}\right)+\textrm{Li}_2 \left(\frac{f}{2r}\right)-\textrm{Li}_2 \left(-\frac{f}{8}\right)-\textrm{Li}_2 \left(\frac{g}{2r}\right)}\right.\nonumber \\
&&\left.{-\textrm{Li}_2 \left(\frac{c^2}{4ac-h}\right)}\right\}. \label{MI11}
\end{eqnarray}

\acknowledgments
This work is supported by the Natural Science Foundation of China under the Grant No. 12065006. \\

\providecommand{\href}[2]{#2}\begingroup\raggedright

\end{document}